\documentclass[a4paper,twocolumn]{article}
\usepackage[dvips]{graphicx}
\begin{document}

\title{Realization of a resonant non-linear phase flip 
in cavity quantum electrodynamics}

\author{Holger F. Hofmann $^{a,b}$, Kunihiro Kojima $^b$, 
Shigeki Takeuchi $^{a,b}$ , and Keiji Sasaki $^b$\\
$^a$ PRESTO, Japan Science and
Technology Corporation (JST)\\  
$^b$Research Institute for Electronic Science,\\ 
Hokkaido University, Sapporo 060-0812\\
Tel/Fax: 011-706-2648\\ 
e-mail: h.hofmann@osa.org}

\date{}

\maketitle

\begin{abstract}
Optical nonlinearities sensitive to individual photons
may be extremely useful as elements in quantum logic
circuits for photonic qubits. A much cited example is
the work of Turchette et al. [Phys. Rev. Lett. 75, 4710 
(1995)], in which a phase shift of about 10 degrees was 
reported. To improve this result, we propose a single
sided cavity geometry with minimal cavity losses. 
It should then be possible to achieve a nonlinear
phase shift of 180 degrees.
\\[0.2cm]
Keywords: quantum level nonlinearity, cavity quantum 
electrodynamics\\
\end{abstract}

\section{Introduction}

One of the fundamental challenges in realizing optical
quantum computation and similar applications of quantum
information encoded in photonic qubits is the relative
weakness of nonlinear interactions that could couple 
individual photons. Among the many proposals made to
overcome this limitation, some of the most impressive
results have been obtained in cavity quantum 
electrodynamics \cite{Nie}. In particular, the experimental
observation of a nonlinear phase shift of 10 degrees reported
by Turchette et al. has attracted a lot of attention
\cite{Tur95}. The mechanism by which this phase shift is
obtained is the saturation of a single two level atom 
confined in a single mode cavity. Therefore, the phase shift
depends on a non-vanishing excitation of the atom and
the results obtained for the magnitude of the phase shift
should increase as the pump light approaches resonance.
The main limitation of this resonant improvement of the
nonlinear phase shift is set by the sensitivity of the
cavity transmission and the transversal losses to the
atomic resonance. We therefore propose a reflection 
geometry with negligible transversal losses as an optimized
device for obtaining a nonlinear phase shift sensitive
to individual photons \cite{Hof02}. With such a device,
it should be possible to achieve nonlinear phase shifts of 
180 degrees. 

\section{Optical confinement and the purpose of a bad cavity}

The source of the strong nonlinearity observed in cavity
quantum electrodynamics is the two level atom that changes
its optical properties drastically in its transition from
ground state to excited state. In principle, this nonlinear 
change of the optical response is found in any individual
two level system. However, it is difficult to isolate a single
two level atom in such a way that the optical input and output
fields can be focussed efficiently on this individual 
atomic system.
For this purpose, it is convenient to use a resonant cavity with
a length equal to half a wavelength. In such a cavity, it is
possible to suppress the coupling of the atom with any mode other
than the single resonant cavity mode. As a result, all of the 
light entering the cavity will interact with the atom, and most of
the light radiated by the atom will be emitted through the cavity
mode. Effectively, it is the purpose of the cavity to focus all
the input light on the atom and to guide all the output light to
the detectors. The longitudinal confinement given by the cavity
lifetime is not relevant for this purpose. It is therefore 
possible to achieve very good results in the bad cavity regime,
where the cavity decay rate $\kappa$ is much faster than
the coupling rate $g$ of the field-atom interaction.

\begin{figure}
\setlength{\unitlength}{0.8pt}
\begin{picture}(300,200)
%\put(0,0){\framebox(300,200){}}
\bezier{400}(240,60)(260,100)(240,140)
\bezier{400}(250,60)(270,100)(250,140)
\put(240,60){\line(1,0){10}}
\put(240,140){\line(1,0){10}}
\bezier{400}(180,60)(160,100)(180,140)
\bezier{400}(175,60)(155,100)(175,140)
\put(175,60){\line(1,0){5}}
\put(175,140){\line(1,0){5}}
\put(235,30){\makebox(20,20){$R_B \approx 1$}}
\put(167,30){\makebox(20,20){$R_F \ll 1$}}

\put(190,77){\line(1,0){40}}
\put(195,60){\makebox(30,15){$\mid \! G \rangle$}}
\put(190,123){\line(1,0){40}}
\put(195,125){\makebox(30,15){$\mid \! E \rangle$}}

\put(210,80){\line(0,1){40}}
\put(210,80){\line(1,2){8}}
\put(210,80){\line(-1,2){8}}
\put(210,120){\line(1,-2){8}}
\put(210,120){\line(-1,-2){8}}

\put(20,90){\line(1,0){140}}
\put(20,110){\line(1,0){140}}
\put(160,110){\line(-2,1){15}}
\put(160,110){\line(-2,-1){15}}
\put(20,90){\line(2,1){15}}
\put(20,90){\line(2,-1){15}}
\put(70,110){\makebox(30,20){$b_{\mbox{in}}$}}
\put(70,70){\makebox(50,20){$b_{\mbox{out}}$}}
\end{picture}
\caption{\label{geometry} Illustration of the dissipation 
free realization of an atomic nonlinearity.
$R_F$ and $R_B$ denote the reflectivities of 
the two mirrors, $b_{\mbox{in}}$ and $b_{\mbox{out}}$
represent the external field interacting with the atom
through the cavity.}
\end{figure}
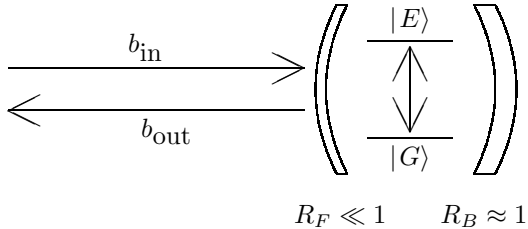

Figure \ref{geometry} illustrates a one sided cavity setup
with negligible transversal losses ($\gamma \ll \kappa, g$).
In this setup, the two level atom inside the cavity only
interacts with the one dimensional fields on the left
side of the cavity. These fields, represented by 
the input $b_{\mbox{\,in}}$ and the output $b_{\mbox{out}}$,
have a transversal profile defined by the cavity mode.
If the input light and the detectors match this 
mode profile, maximal interaction with the two level
atom is obtained.

The conventional cavity parameters characterizing the
field-atom interaction are the cavity damping rate
$\kappa$, the atom-field coupling rate $g$, and the
loss rate $\gamma$ describing the spontaneous emission
into non-cavity modes. However, in the bad cavity regime,
the field dynamics can be adiabatically eliminated.
The cavity damping rate and the atom-field coupling
then define the rate of emission through the cavity
as $2 g^2/\kappa$. A bad cavity therefore divides the
total spontaneous emission of the atom into two parts,
one guided by the cavity into a well focussed beam,
the other spread out over all directions in three
dimensional space. 

This situation is also reflected by the equations
of motion for the atomic system obtained from the
standard equations of cavity quantum electrodynamics
by adiabatic elimination of the cavity field.
Due to the adiabatic elimination, these equations
are now equivalent to the Bloch equations of a
single two level atom,
\begin{eqnarray}
\label{eq:bloch}
\frac{d}{dt} \langle \hat{\sigma}_- \rangle
&=& -\Gamma \langle \hat{\sigma}_- \rangle + 
2 \sqrt{2 \beta \Gamma}\; b_{\mbox{in}}(t)\; 
\langle \hat{\sigma}_z \rangle
\nonumber \\
\frac{d}{dt} \langle \hat{\sigma}_z \rangle
&=& - 2 \Gamma \left(\langle \hat{\sigma}_z \rangle 
                     + \frac{1}{2}\right) 
\nonumber \\&& \hspace{-1cm}
- 
\sqrt{2 \beta \Gamma}\left(\; b^*_{\mbox{in}}(t)\; 
\langle \hat{\sigma}_- \rangle + \;b_{\mbox{in}}(t)\; 
\langle \hat{\sigma}_- \rangle^*\right).
\end{eqnarray}
In this formulation, the atom dynamics itself depends
only on the dipole relaxation rate $\Gamma$ given by
\begin{equation}
\Gamma = \frac{g^2}{\kappa} + \frac{\gamma}{2}.
\end{equation}
Since this dipole relaxation rate is entirely due
to spontaneous emission, the total spontaneous
emission rate is equal to $2 \Gamma$ and the 
two contributions to $\Gamma$ can be identified
with spontaneous emission through the cavity 
($g^2/\kappa$) and with emission into non-cavity
modes ($\gamma/2$), respectively.
The only other coefficient describing the cavity
geometry is the spontaneous emission factor $\beta$
that describes the fraction of the total spontaneous
emission emitted through the cavity, such that the
coupling between the atom and the cavity input field 
$b_{\mbox{in}}$ is given by
\begin{equation}
\beta \Gamma = \frac{g^2}{\kappa}.
\end{equation}
Note that the input field $b_{\mbox{\,in}}$ is normalized
in such a way that $|b_{\mbox{\,in}}|^2$ is equal to the
photon current in the coherent input beam.

With the coefficients given above, it is also possible to
determine the output field. According to input-output
theory \cite{Wal}, it is given by a sum representing 
interference between the reflected light and the dipole
emission,
\begin{equation}
\label{eq:inout}
b_{\mbox{out}}(t) = b_{\mbox{in}}(t) + \sqrt{2\beta\Gamma} 
\langle \hat{\sigma}_- \rangle.
\end{equation}
Again, the spontaneous emission factor $\beta$ modifies 
the coupling between the atom and the field. It is 
therefore interesting to investigate how a variation
of $\beta$ affects the nonlinear response of the atom.

\begin{figure*}[t]
\setlength{\unitlength}{0.9pt}
\begin{picture}(480,350)
\put(40,325){\makebox(200,20){(a) Phase shift for 
$\beta=0.2$}}
\put(40,190){\makebox(200,150){\includegraphics
[width=6cm]{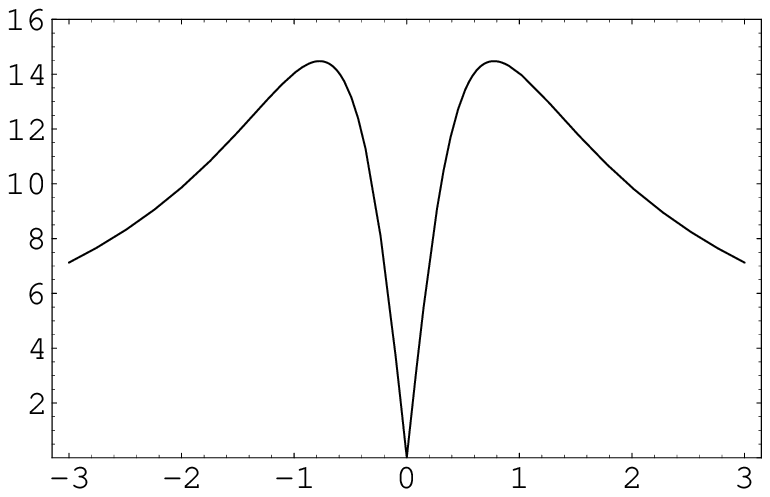}}}
\put(15,270){\makebox(30,10){\large $|\Delta \phi|$}}
\put(15,255){\makebox(30,10){(deg.)}}
\put(132,185){\makebox(40,10){\large $\omega/\Gamma$}}

\put(280,325){\makebox(200,20){(b) Phase shift for 
$\beta=0.4$}}
\put(280,190){\makebox(200,150){\includegraphics[width=6cm]
{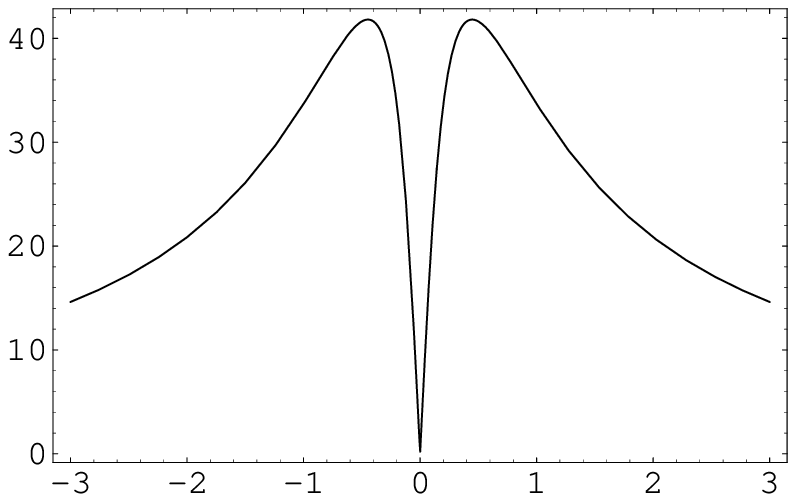}}}
\put(255,270){\makebox(30,10){\large $|\Delta \phi|$}}
\put(255,255){\makebox(30,10){(deg.)}}
\put(372,185){\makebox(40,10){\large $\omega/\Gamma$}}

\put(40,145){\makebox(200,20){(c) Phase shift for 
$\beta=0.6$}}
\put(40,10){\makebox(200,150){\includegraphics[width=6cm]{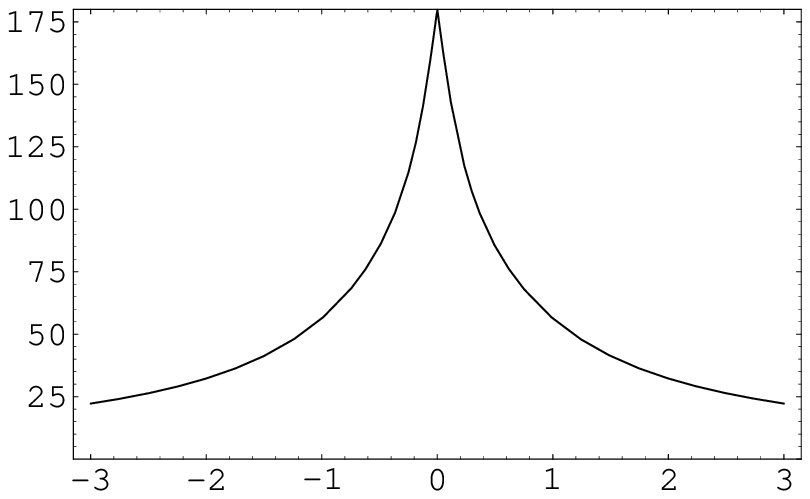}}}
\put(15,90){\makebox(30,10){\large $|\Delta\phi|$}}
\put(15,75){\makebox(30,10){(deg.)}}
\put(132,5){\makebox(40,10){\large $\omega/\Gamma$}}

\put(280,145){\makebox(200,20){(d) Phase shift for $\beta=0.8$}}
\put(280,10){\makebox(200,150){\includegraphics[width=6cm]{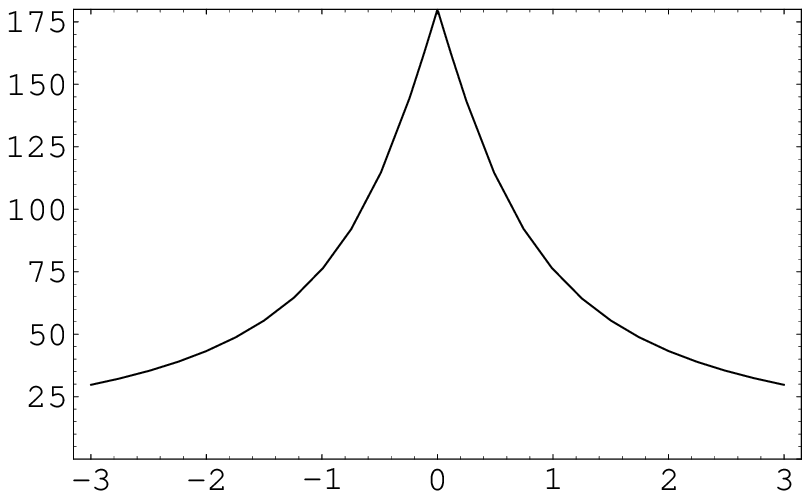}}}
\put(255,90){\makebox(30,10){\large $|\Delta\phi|$}}
\put(255,75){\makebox(30,10){(deg.)}}
\put(372,5){\makebox(40,10){\large $\omega/\Gamma$}}

\end{picture}
\setlength{\unitlength}{1pt}
\caption{\label{beta} Dependence of the nonlinear phase 
shift $\Delta \phi$ on the detuning frequency $\omega$
for different spontaneous emission factors $\beta$.}
\end{figure*}

\section{Nonlinear response and phase shift optimization}

For a continuous input field of frequency $\omega$ and
amplitude $b_{\mbox{\,in}}$, the steady state solution of
equation (\ref{eq:bloch}) is given by
\begin{eqnarray}
\label{eq:stat}
\langle \hat{\sigma}_z \rangle &=& -
\frac{\Gamma^2+\omega^2}{2 (\Gamma^2+\omega^2 
+4 \beta \Gamma |b_{\mbox{in}}|^2)} 
\nonumber \\
\langle \hat{\sigma}_- \rangle &=& -
\frac{\sqrt{2 \beta\Gamma}(\Gamma-i\omega)}
{\Gamma^2+\omega^2 +4 \beta \Gamma |b_{\mbox{in}}|^2} 
\; b_{\mbox{in}}.
\end{eqnarray}
The output field reflected from the cavity is then 
obtained from equation (\ref{eq:inout}). It reads
\begin{eqnarray}
b_{\mbox{out}} &=&
\left( 
1- \frac{2\beta \Gamma (\Gamma-i\omega)}
{\Gamma^2+\omega^2 +4 \beta \Gamma |b_{\mbox{in}}|^2} 
\right)\; b_{\mbox{in}}
\nonumber \\[0.2cm]
&=& 
\frac{((1-2\beta)\Gamma+i\omega)(\Gamma-i\omega)
+4 \beta \Gamma |b_{\mbox{in}}|^2}
{\Gamma^2+\omega^2 +4 \beta \Gamma |b_{\mbox{in}}|^2}
\; b_{\mbox{in}}
.
\nonumber \\
\end{eqnarray}
This nonlinear response function characterizes the transition
from weak fields to strong fields. Close to resonance, these
regimes can be defined with respect to the dipole relaxation
rate as $|b_{\mbox{in}}|^2 \ll \Gamma/\beta$ and 
$|b_{\mbox{in}}|^2 \gg \Gamma/\beta$, respectively. 
In other words, $\Gamma/\beta$ defines the intensity of the
input light at which the nonlinear effect becomes significant.

In the limit of high intensity input fields, 
$|b_{\mbox{in}}|^2 \gg \Gamma/\beta$, the atom is completely
saturated ($\langle \hat{\sigma}_z \rangle = 0$), and the 
response is that of 100\% reflection at an empty cavity,
\begin{equation}
\label{eq:strong}
b_{\mbox{out}} \approx b_{\mbox{in}}.
\end{equation}
The nonlinear phase shift obtained by this saturation effect
is therefore equal to the phase shift caused by the single
atom in the linear limit of weak fields, 
$|b_{\mbox{in}}|^2 \ll \Gamma/\beta$). In this case,
the response of the atomic system is given by
\begin{equation}
\label{eq:weak}
b_{\mbox{out}} \approx 
\frac{\Gamma(1-2\beta)+i\omega}
{\Gamma+i\omega} 
\;
b_{\mbox{in}}.
\end{equation}
The phase shift caused by this linear limit of the
response function can then be defined by
\begin{eqnarray}
|\Delta \phi| &=&
\arccos\left(\frac{\Gamma(1-2\beta)}
{\sqrt{\Gamma^2(1-2\beta)^2+\omega^2}}\right)
\nonumber \\[0.2cm] &&
-\arccos\left(\frac{\Gamma}{\sqrt{\Gamma^2+\omega^2}}\right).
\end{eqnarray}
As noted above, this phase shift then describes the total 
nonlinear phase shift that can be obtained by varying
the input intensity.

Figure \ref{beta} shows the spectra of nonlinear phase shifts 
obtained for various spontaneous emission factors $\beta$.
Note that the maximal phase shifts obtained off resonance
are significantly smaller than 180 degrees. Specifically, 
the maximal phase shift at $\beta=0.2$ is about 14.5 degrees
at $\omega=0.78 \Gamma$,
and the maximal phase shift at $\beta=0.4$ is about 42 degrees
at $\omega=0.45 \Gamma$. For both $\beta=0.6$ and $\beta=0.8$,
the phase shift reaches its maximum value of 180 degrees at
$\omega=0$. The increase from $\beta=0.6$ to $\beta=0.8$ 
only increases the off-resonant phase shifts.

\section{Amplitude nonlinearity and switching efficiency}
The nonlinear change from the weak field response 
described by equation (\ref{eq:weak}) to the strong field
response given by equation (\ref{eq:strong}) also includes
an increase of the reflected amplitude.
For weak fields, only a fraction $\eta$ of the input
intensity is reflected. This fraction is given by
\begin{equation}
\label{eq:eta}
\eta =
\frac{\Gamma^2(1-2\beta)^2+\omega^2}
{\Gamma^2+\omega^2}.
\end{equation}
For strong fields, all of the input light is reflected.
The amplitude nonlinearity is therefore characterized 
by a change of reflectivity from $\eta$ to one. In order to 
realize a loss free phase nonlinearity, it is desireable
to improve the efficiency $\eta$ to be close to one.
This can be achieved by varying the frequency or by
varying the spontaneous emission factor $\beta$.
However, the previous discussion has shown that high
phase shifts can only be achieved close to resonance
at high $\beta$. The only way to improve efficiency
in this regime is to improve the value of $\beta$
by avoiding spontaneous emission into non-cavity 
modes. 

At resonance, the efficiency is given by 
$\eta=(1-2\beta)^2$. For the values of $\beta$ shown 
in figure \ref{beta}, the efficiencies at resonance
are therefore equal to $0.04$ for $\beta=0.6$,
and equal to $0.36$ for $\beta=0.8$. 
The suppression of spontaneous emission to non-cavity 
modes is therefore important for the realization of
a maximally sensitive optical phase switch.

\section{Conclusions}
In order to realize a nonlinear phase shift of 180 degrees
using the optical properties of a single atom in a cavity,
it is necessary to reduce the fraction of spontaneous 
emission emitted into non-cavity modes below 
$1-\beta=0.5$. It is then possible to obtain the desired
phase shift at resonance with the atom. However, losses
make a further improvement towards $\beta=1$ desireable.
In principle, a completely loss free nonlinear phase
flip of 180 degrees can be achieved if all the light
emitted from the atom is confined by the cavity at $\beta=1$
\cite{Hof02}.

\section*{Acknowledgements}
Part of this work was supported by the program "Research and
Development on Quantum Communication Technology" of the
Ministry of Public Management, Home Affairs, Posts and
Telecommunications of Japan.

%=========================================================

%=========================================================

\end{document}